\documentstyle[prb,preprint,aps]{revtex}
\begin{document}
\draft
%
%
\title{Nonlinear Magneto-Optical Response
of $s$- and $d$-Wave Superconductors}
\author{ J. Schmalian and W. H\"ubner}
\address{Institut f\"ur Theoretische Physik,
Freie Universit\"at Berlin, Arnimallee 14,
14195 Berlin , Germany}
\date{\today}
\maketitle
\begin{abstract}
The  nonlinear magneto-optical response of $s$-
and $d$-wave superconductors is
discussed. We carry out the symmetry analysis
of the nonlinear
magneto-optical susceptibility in the
superconducting
state. Due to the
surface sensitivity of the nonlinear optical
response
 for
systems with bulk inversion symmetry, we
 perform a
group theoretical
classification of the superconducting order
parameter close to a surface.
For the first time, the mixing of singlet and
triplet pairing states induced
by spin-orbit coupling is systematically
 taken into account.
We show that the interference of singlet
 and triplet pairing
states leads to an
observable contribution of the nonlinear
 magneto-optical Kerr
 effect.
This effect is not only sensitive to the
 anisotropy of the
gap function but
also to the symmetry itself.
In view of the current
discussion of the order
parameter symmetry of High-T$_c$
 superconductors,
results for a tetragonal
system with bulk singlet pairing  for
various pairing
symmetries are discussed.
\end{abstract}
\pacs{74.25.Gz,74.25.-q,74.25.Nf,78.20.Ls}
\newpage
%
\section{  INTRODUCTION}
The investigation of the symmetry of the superconducting
order parameters
is  currently one of the most exciting  problems in the
field of
high-$T_c$
research.~\cite{MBK92,SDW93,HBM93,KCN94,DES94,WVL93,CSY94,BO94,TKC94}
Many experiments suggest that the order
parameter is
anisotropic.~\cite{MBK92,SDW93,HBM93,KCN94,DES94}
Furthermore, the observation of $\pi$ phase shifts
in corner junctions
\cite{WVL93,BO94,TKC94}is consistent with a
$d_{x^2-y^2}$
symmetry of the gap function, although another
experiment~\cite{CSY94}
favors a more conventional $s$-wave pairing.
On the other hand, the
occurrence of a finite tunneling current perpendicular
to the planes
\cite{SGM94} seems to be incompatible with $d$-wave
symmetry in a simple
picture.
In view of this debate, it is important to develop
alternative
experimental techniques which are able to discriminate
between the
various symmetries of the gap function.
In particular, it is interesting to measure the
symmetry of the gap function
without the necessity of a tunneling contact, which
always
includes the problems of a residual magnetic field,
trapped flux in the
tunneling circuit, or singularities in the supercurrent
flow at the
corner~\cite{VH95}.

On the other hand, optical second harmonic generation
has widely been used
as a probe for two dimensional optical, electronic and
structural properties  at various interfaces such as
solid-vacuum,
solid-gas, solid-solid, solid-liquid, or
liquid-gas~\cite{shen,richmond}.
Recently, the
nonlinear magneto-optical Kerr-effect has become a
new nonlinear optical
method for the investigation of low-dimensional magnetic
structures~\cite{pan,hub2,hub3}.
This effect is interface sensitive within the electric
dipole-approximation
and directly probes, unlike methods based on Raman
scattering, low-energy
magnetic excitations ($\sim$ meV) using optical
interband transitions
($\sim$ eV). In contrast to linear optical techniques,
the nonlinear
magneto-optical Kerr effect yields an excellent signal
to noise ratio,
since it is free of linear nonmagnetic background
radiation. This
advantage manifests itself both in the large magnetic
intensity
contrast upon magnetization inversion,
\[\frac{I({\bf M})-I(-{\bf M})}{I({\bf M})+I(-{\bf M})}
\]
of about 40 \% for Fe surfaces\cite{reif,hub4} compared
to a linear effect of
typically 0.1 \% and in a giant nonlinear Kerr rotation
$\phi_{K}^{(2)}$ which,
depending on the chosen Kerr-configuration, can be
tuned up to 90$^{\circ}$
nearly at will~\cite{pusto,kirschner,theo}.
This corresponds to an enhancement by 2 - 3 orders of
 magnitude compared to
the usual linear Kerr rotation, which is even reduced
in thin films.
Thus, it allows for the determination of the magnetic
interface symmetry,
including the magnetic ``easy axis''~\cite{hub5}.

It is well-known that the symmetry of the nonlinear
optical susceptibility is
strongly affected by a magnetization or an external
applied magnetic
field.~\cite{pan} Thus it is of considerable interest
to extend the symmetry
analysis of the nonlinear magneto-optical response to
the superconducting state
for different symmetries of the superconducting order
 parameter.
Therefore, we propose in this paper a new theory for
 the nonlinear
optical response of a superconductor in the presence
 of a magnetic field,
which is able to distinguish
between certain symmetries of the gap function and
might stimulate
corresponding experiments.
In the following, we present our theory and show that
 indeed
a symmetry dependent contribution to the magneto-optical
 response without
tunneling contact results in optical second harmonic
 generation.
Although it is, due to the gauge invariance,
impossible to measure
the phase of the superconducting order parameter
 without tunneling
contact, it is still possible to measure its
symmetry, not solely
its magnitude.
This results from the interference of different pairing
amplitudes in the dipole
matrix elements of the {\em three} transitions in
nonlinear optics.
Due to the surface sensitivity of nonlinear optics
 for systems with bulk
inversion symmetry, one can take advantage of the
broken inversion symmetry at the surface.
This is of interest, because Cooper pairing,
together with the
always present spin-orbit coupling, is then no
more purely
singlet- or triplet-like.
The interference of the singlet and triplet pairing
states, which is linear in
spin-orbit interaction, leads to the symmetry sensitive
contribution of the nonlinear optical response for
systems in an external magnetic field.
Note that the possible importance of mixed singlet
and triplet pairing states at
the interface of a tunneling contact between a
heavy fermion and a conventional
superconductor has already  been  discussed by
 Fenton\cite{F85}.
In order to obtain a more systematic insight
 into these phenomena,
we perform the description of a superconductor at
the surface of a bulk inversion symmetric system,
including spin-orbit
interaction and give a group-theoretical classification
of the irreducible
representations of the gap function.
This is of importance for a detailed calculation of
the tensor elements of the
nonlinear optical susceptibility which is performed
in the second part
of the theory.
Finally, we present our results concerning the
symmetry dependence of the
corresponding experiment and discuss in detail
 the differences
of the line shapes of the optical spectra.
We find that it is possible to discriminate between
an isotropic $s$-wave,
a $s_{x ^2 + y ^2}$- or  $d_{x ^2 - y ^2}$- wave
and a $d_{x  y }$- wave.
However, no symmetry dependent differences between
the  $s_{x ^2 + y ^2}$- and
the $d_{x ^2 - y ^2}$- waves occur, which are
actually the most discussed
symmetries of the high-T$_c$ systems.
Nevertheless nonlinear magneto-optics is shown
to be an
alternative and complementary method to gain
insight into the
symmetry of the superconducting gap function.

\section{ THEORY}
The strategy of this section is to calculate the
effects of $s$- and $d$-wave
superconductivity on the nonlinear optical
susceptibility tensor $\chi^{(2)}$,
which are due to (a) the modification of the
bandstructure and thus the
resonance denominators and (b) due to the
symmetry of the superconducting
order-parameter affecting the wave functions
of the optical electric-dipole
transition matrix elements. Thus, we proceed
as follows: (i) First we set up
the superconducting BCS-type Hamiltonian and
perform its group theoretical
analysis. In order to obtain the desired
sensitivity of nonlinear optics
to the symmetry of the gap function this
requires, as it will turn out,
the simultaneous absence of inversion symmetry,
presence of an
external magnetic field breaking time reversal,
and spin-orbit interaction
coupling singlet and triplet pairs. (ii) Making
use of this symmetry
classification and the mentioned constraints we
then calculate the nonlinear
magneto-optical response in $s$- and $d$- wave
superconductors from the
appropriate current-current-current correlation
function and propose
a suitable experimental geometry for the observation
of this new nonlinear
magneto-optical effect in superconductors.

In our theory for the nonlinear magneto-optical
 response of unconventional
superconductors, the superconducting state is
described within the BCS
theory~\cite{BCS57}. The corresponding Hamiltonian
with arbitrary pairing
symmetry and in a magnetic field $\vec{h}$ is
given by
\begin{equation}
H=\sum_{{\bf k}\mu} \Psi_{{\bf k}\mu}^\dagger
{\cal H}_{{\bf k}\mu}
\Psi_{{\bf k}\mu}\, ,
\end{equation}
with the four-component Nambu spinor
$\Psi_{{\bf k} \mu} = (c_{{\bf k} \mu \uparrow},
c_{{\bf k} \mu \downarrow},
       c^\dagger_{-{\bf k} \mu \uparrow},
c^\dagger_{-{\bf k} \mu \downarrow})$.
Here, $c^\dagger_{{\bf k} \mu \sigma}$ is the
creation operator of an electron
with momentum ${\bf k}$, band index $\mu$ and
spin $\sigma$.
The $(4 \times 4)$ matrix  ${\cal H}_{{\bf k}\mu}$
can be expressed in terms
of $(2 \times 2)$ block matrices:
\begin{equation}
{\cal H}_{ {\bf k} \mu }= \left( \begin{array}{cc}
           \varepsilon_{{\bf k} \mu}
           \hat{\sigma}^o - \vec{h} \cdot
           \hat{\vec{\sigma}} &
            \hat{\Delta}_{{\bf k}\mu} \\
           -\hat{\Delta}^\ast_{-{\bf k}\mu} &\
           -\varepsilon_{{\bf k}\mu}
           \hat{\sigma}^o  +\vec{h}\cdot
           \hat{\vec{\sigma}}^\ast
           \end{array} \right) \, .
\label{HBCS}
\end{equation}
The block matrices are expanded in terms of
 the unit matrix $\hat{\sigma}^o$
and the vector of the Pauli matrices
$\hat{\vec{\sigma}}$.
This notation is close to that of Sigrist
and Ueda~\cite{SU91}.
The symmetry of the superconducting order
parameter is characterized by the
gap function
 $\Delta_{\sigma \sigma ' ;{\bf k}\mu}=
\langle c_{{\bf k} \sigma \mu}
c_{-{\bf k} \sigma ' \mu } \rangle$.
  We neglect any diamagnetic, i.e. Meissner
effect of the magnetic field,
but assume a large penetration depth at the
 surface and no influence of
the vortex structure to the optical spectrum.
This seems to be reasonable at least for the
excitations in the interband
regime discussed in this paper.
$\Delta_{\sigma \sigma' ;{\bf k}\mu}$ is
decomposed in the usual way in singlet states
($\Delta ^o_{{\bf k} \mu}=\Delta ^o_{-{\bf k}\mu}$)
and triplet states
($\vec{d}_{{\bf k}\mu}=-\vec{d}_{-{\bf k}\mu}$):
\begin{equation}
 \hat{\Delta}_{{\bf k}\mu}=\left(\Delta ^o_{{\bf k}\mu}
\hat{\sigma}^o +
\vec{d}_{{\bf k}\mu} \cdot \hat{\vec{\sigma}} \right)
i\hat{\sigma} ^y \, .
\label{gapexp}
\end{equation}
The symmetry of $\Delta ^o_{{\bf k}\mu}$ and
 $\vec{d}_{{\bf k}\mu}$
with respect to the
transition from ${\bf k}$ to $-{\bf k}$ is a
direct consequence of the
Pauli principle.
Since we  consider the states at a surface, ${\bf k}$
refers to the
two dimensional in-plane momentum.

Below the transition temperature $T_c$, the symmetry
of a system is reduced
compared to the high temperature phase.
The symmetry group $G$ of the high temperature phase
 is determined by the
symmetry operations which keep the Hamiltonian for
 $ \hat{\Delta}_{\bf k}=0$
invariant.
We consider a system which is  for $\vec{h}=0$
invariant with respect to the group
\begin{equation}
G=g \times K \times U(1) \, ,
\end{equation}
where $g$, $K$ and $U(1)$ are the point group,
time reversal operation and the
gauge group of multiplication of electron creation
operator by an arbitrary
phase, respectively.~\cite{SU91,A85,VG85,B85,SR89,SR92}.
In the ordinary case the normal-state gauge
symmetry is broken
at the superconducting phase transition, i.e.
the residual symmetry group is
$g \times K$. This is called a conventional
superconductor.
In unconventional superconductors however,
the symmetry is lower than
$g \times K$. At the transition temperature,
the BCS - gap equation is an
eigenvalue equation and consequently, an eigenvector
$\Delta_{\sigma \sigma' ;{\bf k} \mu }$  belongs to
one of the  irreducible representations
${\cal D}$ of the group  $G$.
If ${\cal D}$ is the unit representation
${\cal A}_1$ (or ${\cal A}_{1g}$
for systems with inversion symmetry) conventional
superconductivity occurs.
In all other cases the superconductivity is
unconventional.
In order to discuss the various symmetry states
of the order parameter,
one has to generate all irreducible representations
of the gap function,
where, due to spin-orbit coupling, the spin
degrees of freedom cannot be
transformed independently from the spatial
(orbital) coordinates.
For various point groups this symmetry
classification has been
performed~\cite{SU91,A85,VG85,B85,SR89,SR92}.
In all these cases the inversion
operation $C_{i}$ is an element of the group $G$.
Since we are interested in the investigation of
superconducting properties with
 surface sensitive nonlinear optical experiments,
 we have to take the effect of
broken inversion symmetry into account.
In order to be specific, we consider the surface
of a tetragonal system
(bulk point group D$_{4 h}$) with residual point
group $C_{4v}$.
This group has five irreducible representations:
four  of  dimension one
(${\cal A}_1$, ${\cal A}_2$, ${\cal B}_1$ and
${\cal B}_2$) and one of
dimension two (${\cal E}$).
The isotropic  $s_o$ and and anisotropic
$s_{x^2+y^2}\; $s-waves transform
as ${\cal A}_1$, the $d_{x^2-y^2}$-wave as
${\cal B}_1$,
the $d_{xy}$-wave as ${\cal B}_2$, a
$d_{x^2-y^2}d_{xy}$-wave
as ${\cal A}_2$ and the $p_x$,  $p_y$ waves
as ${\cal E}$.

In order to classify the irreducible
representations of the gap function, we
have to analyze the transformation properties of
$\Delta_{\sigma \sigma' ;{\bf k}\mu}$.
Applying an element $R$ of the point group to
$\hat{\Delta}_{{\bf k}\mu}$,
the following transformation of the singlet and
triplet part results:
\begin{equation}
R\, \hat{\Delta}_{{\bf k}\mu}= \left(
\Delta ^o_{{\cal D}^{(1)}_R{\bf k} \mu} +
\tilde{{\cal D}}^{(1)}_R
\vec{d}_{{\cal D}^{(1)}_R{\bf k}\mu}
\cdot \hat{\vec{\sigma}} \right) i\hat{\sigma ^y} \, .
\end{equation}
Here, ${\cal D}^{(1)}_R$ is the representation
of $R \in G$ which transforms the
coordinates.
If one considers the transformation of a Pauli
spinor with respect to a
combination $R=R_o C_i$ of the  inversion
operation  $C_i$  and a rotation
$R_o$, only the rotational part has to be
applied to the spinor, i.e. the
representation of $R$ in spin space is
${\cal D}^{(1/2)}_{R_o}$.
Consequently, for the vector in spin space
$\vec{d}_{\bf k}$, the representation
$\tilde{{\cal D}}^{(1)}_R \equiv {\cal D}^{(1)}_{R_o}$,
 where the inversion
operation is replaced by the identical
transformation, has to be applied.
Therefore, one finds in the {\em bulk} of a
system with inversion symmetry:
$ C_i\, \hat{\Delta}_{{\bf k}\mu}=
\left(\Delta ^o_{{\bf k}\mu} -
  \vec{d}_{{\bf k}\mu}
\cdot \hat{\vec{\sigma}} \right) i\hat{\sigma ^y}$,
since the vector $\vec{d}$ is not affected
directly by the inversion operation.
The minus sign results from the inversion
of ${\bf k}$ to $-{\bf k}$.
Consequently,  in the {\em bulk} , the singlet
 and triplet part belong to
different irreducible representations and
either singlet or triplet
superconductivity occurs.
In distinction to this, a coexistence of singlet
and triplet pairing states is
possible for systems without inversion symmetry,
i.e. at the {\em surface}.
Now, the in-plane inversion operation  can
be realized by a rotation
(rotation by $\pi$ with $z$-axis as rotation axis).
This rotation transforms the vector $\vec{d}$
to $-\vec{d}$ and
the minus sign of the transformation
${\bf k} \rightarrow -{\bf k}$
is canceled. Consequently, the irreducible
representations of the gap function
at the  {\em surface} contain both, singlet
and triplet parts.

{}From these considerations one obtains the
irreducible representations of
the gap function from the simultaneous
Clebsch-Gordon coupling of orbital
and spin degrees of freedom.
The results for the simultaneously occurring
singlet and triplet
part of the pairing amplitude are given in
the table.
Sigrist and Rice~\cite{SR92} calculated the
 irreducible
representations of the tetragonal group
$D_{4h}$. Since
$D_{4h}=C_{4v} \times C_i$, it is
straightforward to check the above results
by reducing the subduced representations of $D_{4h}$.
One finds: ${\cal A}_1={\cal A}_{1g} \oplus {\cal A}_{2u}$,
${\cal A}_2={\cal A}_{2g} \oplus {\cal A}_{1u}$,
${\cal B}_1={\cal B}_{1g} \oplus {\cal B}_{2u}$,
${\cal B}_2={\cal B}_{2g} \oplus {\cal B}_{1u}$ and
${\cal E}={\cal E}_{g} \oplus {\cal E}_{u}$ which
leads to the results of the
table. Here, $g$ and $u$ refer to the  irreducible
representations of $D_{4h}$
with even and odd parity.
The irreducible representations of table~\ref{tab1}
are the possible symmetry
states of a superconductor on the surface of a
tetragonal system and with spin
orbit interaction. If one neglects the spin orbit
coupling, the spin and orbital
degrees of freedom transform separately and the
singlet and triplet pairing
states decouple again. Therefore, for a bulk
singlet superconductor, the
simultaneously occurring triplet part at the
surface is of the
order of the spin-orbit interaction, and vice
versa.
In heavy fermion systems the spin-orbit interaction
 is large, but even in
transition metals one expects this quantity to
be of the order of 50 meV which,
although being smaller, has nevertheless dramatic
consequences such as the
reorientation of the magnetic easy axis in thin
 ferromagnetic films upon
the increase of the film thickness or the rise
of the temperature.
Finally, we expect the spin-orbit induced triplet
 part to be observable if one
considers a surface sensitive experiment such as
second harmonic generation.

Based on these group theoretical classifications,
we calculate now
the nonlinear magneto-optical susceptibility tensor
of a superconductor
and focus on the interference of the simultaneously
occurring singlet and
triplet part of the gap function.
The optical response in second harmonic generation can
be obtained from the
nonlinear current-current-current correlation function:
\begin{eqnarray}
\chi_{\alpha \beta \gamma}({\bf q}, \omega)    &=&
 \int_{-\infty}^{\infty} \frac{d \epsilon}{\pi}
\int_{-\infty}^{\infty} \frac{d \epsilon '}{\pi}
\int_{-\infty}^{\infty} \frac{d \epsilon ''}{\pi}
I_{\alpha \beta \gamma}({\bf q},\epsilon,\epsilon ',
\epsilon '') \;\times
\nonumber \\
& &
\frac{
\frac{f(\epsilon '')-f(\epsilon ')}
 {\omega +i\delta -\epsilon '' +\epsilon '} \, - \,
\frac{f(\epsilon' )-f(\epsilon )}
                    {\omega +i\delta -\epsilon '
+\epsilon}}
{2(\omega +i\delta) -\epsilon '' +\epsilon }
\end{eqnarray}
where $f(\epsilon)$ is the Fermi function and
the spectral
function
$I_{\alpha \beta \gamma}
({\bf q},\epsilon,\epsilon ',\epsilon '') $
is given by
\begin{equation}
I_{\alpha \beta \gamma}
({\bf q},\epsilon,\epsilon ',\epsilon '')
={\rm Tr} \left( J_{-2{\bf q} \alpha}
\varrho(\epsilon)
J_{{\bf q}\beta} \varrho(\epsilon')
J_{{\bf q} \gamma} \varrho(\epsilon'')\right) \, .
\label{nolispec}
\end{equation}
$J_{{\bf q} \alpha}$ is the $\alpha$-th
component of the
current operator
\begin{equation}
\vec{J}_{\bf q}= \sum_{{\bf k} \sigma \mu \nu}
\vec{j}_{{\bf k}\mu \nu}
c^\dagger_{{\bf k}+\frac{{\bf q}}{2} \sigma \mu}
c_{{\bf k}-\frac{{\bf q}}{2} \sigma \nu}\, ,
\end{equation}
and  $\varrho(\epsilon)=-\frac{1}{\pi}{\rm Im}
(\epsilon+i\delta +H)^{-1}$
is the density of states matrix  with
Hamiltonian $H$ of Eq.~\ref{HBCS}.
The trace has to be performed with respect
to all single particle states, i.e.
the momentum (${\bf k}$), band ($\mu, \nu$),
spin ($ \sigma$) and Nambu degrees
of freedom. In the  following, we consider only
interband transitions
$\mu \neq \nu$, and the limit of the dipole
approximation
${\bf q} \rightarrow {\bf 0}$ can be
performed without special care for
plasmonic excitations.

The evaluation of the trace of Eq.~\ref{nolispec}
is straightforward
with the knowledge of the unitary
transformation ${\cal U}_{{\bf k}\mu}$,
which diagonalizes $H$ and $\varrho(\epsilon)$.
In the following we discuss ${\cal U}_{{\bf k}\mu}$
for a bulk singlet
superconductor in the limit
of weak spin-orbit interaction $\lambda_{s.o.}$
and weak external field
because the phase sensitive contributions of
the optical response vanish if either
$\lambda_{s.o.}$ or the magnetic
field is zero.
For weak external magnetic field, the
eigenvalues are given by
\begin{equation}
E_{{\bf k}\mu}=\pm h \pm
{\cal E}_{{\bf k}\mu}\, ,
\label{eigens}
\end{equation}
where
\begin{equation}
{\cal E}_{{\bf k}\mu}=
\sqrt{\varepsilon_{{\bf k}\mu}^2 +\frac{1}{2}
{\rm tr} \left(\hat{\Delta}_{{\bf k}\mu}^\dagger
\hat{\Delta}_{{\bf k}\mu}
\right)}\, .
\end{equation}
$h$ is the absolute mangnitude of $\vec{h}$ and
${\rm tr}$ denotes the trace in
spin space. $\hat{\Delta}_{{\bf k}\mu}$ belongs
to one of the irreducible
representations of table~\ref{tab1}.
Analogously, the unitary transformation is
given by
\begin{equation}
 {\cal U}_{{\bf k}\mu} ={\cal U}^\Delta_{{\bf k}\mu}\,
{\cal U}^h_{{\bf k}\mu}\, ,
\label{transcomb}
\end{equation}
where ${\cal U}^\Delta_{{\bf k}\mu}$ and
${\cal U}^h_{{\bf k}\mu}$
are the transformations which diagonalize
${\cal H}_{{\bf k}\mu}$
for $\vec{h}=0$ and $\vec{d}_{{\bf k}\mu}=0$,
 respectively.
This is correct up to first order in
$\lambda_{s.o.}$ and $\vec{h}$.
The zero-field transformation
${\cal U}^\Delta_{{\bf k}\mu}$
can be expressed in terms of $(2\times2)$
matrices $\hat{u}_{{\bf k}\mu}$
and $\hat{v}_{{\bf k}\mu}$~\cite{SU91}:
\begin{equation}
 {\cal U}^\Delta_{{\bf k}\mu}=
\left( \begin{array}{cc}
\hat{u}_{{\bf k}\mu} &\hat{v}_{{\bf k}\mu} \\
\hat{v}_{-{\bf k}\mu}^\ast &\hat{u}_{-{\bf k}\mu} ^\ast
\end{array} \right) \, ,
\end{equation}
where
\begin{equation}
\hat{u}_{{\bf k}\mu}=\left(
{\cal E}_{{\bf k}\mu}-\varepsilon_{{\bf k}\mu}
\right) /
{\cal E}_{{\bf k}\mu} \,  \hat{\sigma}^o
\end{equation}
and
\begin{equation}
\hat{v}_{{\bf k}\mu}=
-\hat{\Delta}_{{\bf k}\mu}/{\cal E}_{{\bf k}\mu}\, .
\label{vmatrix}
\end{equation}
Similar to the gap function in
Eq.~\ref{gapexp}, these $(2 \times 2)$ matrices
are expanded in Pauli matrices, leading
to a singlet  part $v^o_{{\bf k}\mu}$
and triplet part $\vec{v}_{{\bf k}\mu}$
of $\hat{v}_{{\bf k}\mu}$.
The transformation ${\cal U}^h_{{\bf k}\mu}$
is determined by  a rotation
in spin space
\begin{equation}
{\cal U}^h_{{\bf k}\mu}= \left( \begin{array}{cc}
\exp\left(-i \vec{a}\cdot \vec{\sigma}\right) & 0 \\
0 & \exp\left(i \vec{a}\cdot \vec{\sigma}^\ast \right)
\end{array} \right) \, ,
\end{equation}
with rotation axis $ \vec{e}_a= -( \vec{e}_z
 \times \vec{e}_h)/
| \vec{e}_z \times \vec{e}_h | $ and angle
$\cos(2a)=\vec{e}_z  \cdot\vec{e}_h$
($\vec{a}=a \vec{e}_a$).
$\vec{e}_h$ is the unit vector in the
direction of $\vec{h}$.
Using this transformation, we can
diagonalize the Hamiltonian leading
to the quasiparticle spinor
\begin{equation}
\Psi_{{\bf k}\mu} = {\cal U}_{{\bf k}\mu}
\Phi_{{\bf k}\mu} \, .
\end{equation}
Here the components of  $\Phi_{{\bf k}\mu}$
are the  destruction operators of
the eigenstates  with eigenvalues
given in Eq.(~\ref{eigens}).

In order to express the current operator
in terms of the Nambu spinors, we have
to consider the  behavior of the matrix
element $\vec{j}_{{\bf k}\mu \nu}$
under the simultaneous transformation
${\bf k} \rightarrow -{\bf k} $ and
$(\mu,\nu) \rightarrow (\nu,\mu)$:
\begin{equation}
\vec{j}_{-{\bf k}\mu \nu} = p^{\mu \nu}_{{\bf k}}
\vec{j}_{{\bf k}\nu \mu} \, .
\end{equation}
Although the phase factors $p^{\mu \nu}_{{\bf k}}$
 depend on the choice of the
phase of the Wannier functions, all observable
 quantities like
$\chi_{\alpha \beta \gamma}({\bf q}, \omega) $
are independent of this choice.
Now, the current operator can be expressed
in terms of the Nambu spinors:
\begin{equation}
\vec{J}=\frac{1}{2} \sum_{{\bf k}  \mu  \nu}
\vec{j}_{{\bf k} \mu \nu}
\Psi^\dagger_{{\bf k} \mu}
\left( \begin{array}{cc}
\hat{\sigma}^o & 0 \\
0 & - p^{\mu \nu}_{{\bf k}} \hat{\sigma}^o
\end{array} \right)
\Psi_{{\bf k}  \nu}\, ,
\label{currNam}
\end{equation}
and  we find for the nonlinear  spectral
function of Eq.(~\ref{nolispec}):
\begin{eqnarray}
\lefteqn{ I_{\alpha \beta \gamma}
(\epsilon,\epsilon ',\epsilon '')
=\frac{1}{8} \sum_{{\bf k}  \mu \nu \kappa}
j^\alpha_{{\bf k} \mu \nu}
j^\beta_{{\bf k} \nu \kappa}
j^\gamma_{{\bf k} \kappa \mu}  }
 \nonumber \\
& & \times {\rm Tr}'
\left\{ {\cal M}_{{\bf k}}^{\mu \nu} \,
\varrho_{{\bf k}}^\nu(\epsilon)  \,
{\cal M}_{{\bf k}}^{\nu \kappa} \,
\varrho_{{\bf k}}^\kappa(\epsilon ')\,
{\cal M}_{{\bf k}}^{\kappa \mu} \,
\varrho_{{\bf k}}^\mu(\epsilon '') \right\} \, .
\label{tr1}
\end{eqnarray}
Here, ${\cal M}_{{\bf k} }^{\mu \nu}$ and
$\varrho_{{\bf k}}^\nu(\epsilon)$ are
($4 \times 4$) matrices in  Nambu and
spin space, whereby
\begin{equation}
{\cal M}_{{\bf k} }^{\mu \nu} =
{\cal U}_{{\bf k}\mu}^{\dagger}
\left( \begin{array}{cc}
 \hat{\sigma}^o & 0 \\
 0 & - p^{\mu \nu}_{{\bf k}} \hat{\sigma}^o
             \end{array} \right)
 {\cal U}_{{\bf k} \nu}
\label{MM}
\end{equation}
results from the transformation of
Eq.(~\ref{currNam}) into the quasiparticle
representation, where
$\varrho_{{\bf k}}^\nu(\epsilon)$ is diagonal.
Consequently, the trace  ${\rm Tr}'$
 has to be performed with respect to the
Nambu and spin degrees of freedom.

Nonlinear optics and in particular
nonlinear magneto-optics offers
a unique method to probe low-lying
excitations close to the Fermi-level
with optical photons, since SHG, which
involves three photons,
takes advantage of an additional degree
of freedom that is absent in linear
optics. Thus it allows to use conventional
 monochromatic, intense, and
tunable pulse laser sources in the ps to
 fs regime such as mode-locked
Ti-sapphire lasers. Besides, in nonlinear
interband optics, there is no
collective plasmon background which is
material insensitive and may
severely hamper the interpretation of
linear optical experiments.

Therefore, we restrict ourselves to a
special interband excitation process.
We  consider the   transition from the
initial state $i$ with energy
$E_i \approx -3 {\rm eV}$, below the
 Fermi energy
to the intermediate state $s$ at the
Fermi level (which is the only
superconducting state) and to the
final state $f$
with energy $E_f \approx 3 {\rm eV}$
above $E_{\rm F}$, i.e.
we consider $\mu=f$, $\kappa=s$, and $\nu=i$.
A possible, but not necessary, origin
of the states $i$ and $f$ might be
due to the Mott-Hubbard splitting of
the hybridized Cu 3d$_{x ^2 - y ^2}$
and O 2p$_{x(y)}$ orbitals.
Since the intermediate state is the
only state with superconducting coherence,
we skip the band index of the matrices
$\hat{u}_{{\bf k}}$ and
$\hat{v}_{{\bf k}}$.
Performing finally the traces in
Eq.(~\ref{tr1}), we obtain:
\begin{equation}
\chi_{\alpha \beta \gamma} ( \omega)=
\chi^{(0)}_{\alpha \beta \gamma}( \omega)+
\chi^{(h)}_{\alpha \beta \gamma}( \omega)
+{\cal O}(h^2) \, ,
\end{equation}
where
\begin{equation}
 \chi^{(0)}_{\alpha \beta \gamma}( \omega)
=\sum_{\bf k} \,j^\alpha_{{\bf k} f  i}
\, j^\beta_{{\bf k} i s} \, j^\gamma_{{\bf k} s f}\,
\left( | u ^o_{\bf k}|^2 \,  G_1({\bf k},\omega)
+  | v ^o_{\bf k}|^2 \,  G_2({\bf k},\omega) \right)
\end{equation}
is the zero field susceptibility tensor
 in second harmonic
generation within the superconducting state
which gives already
a contribution without spin-orbit interaction.
The nonvanishing tensor elements for
 $\chi^{(0)}_{\alpha \beta \gamma}( \omega)$
are the same as in the normal state
($\alpha \beta \gamma \in
\{zzz, zxx,zyy,xzx,xxz,yzy,yyz\}$).
This results here from the transformational
properties of the three matrix
elements $ j^\alpha_{{\bf k} f  i}
j^\beta_{{\bf k} i s}  j^\gamma_{{\bf k} s f}$,
which transform as the
corresponding combination of
the coordinates $x_\alpha x_\beta x_\gamma$,
even if a single matrix element
e.g.  $ j^\alpha_{{\bf k} f  i}$ does not
transform like $x_\alpha$.
The functions  $ G_{1(2)}({\bf k},\omega)$
result from the numerous
combinations of  Fermi functions and energy
denominators which occur
by performing  the traces in spin and Nambu
space.
$\chi^{(h)}_{\alpha \beta \gamma}( \omega)$
 is  the new contribution of the magneto-optical
Kerr effect in the
superconducting state.
This can be seen from the symmetry relations
of the magneto-optical
susceptibility
\begin{equation}
\chi^{(h)}_{\alpha \beta \gamma}( \omega)
=\sum_{\bf k} \,j^\alpha_{{\bf k} f  i}
\, j^\beta_{{\bf k} i s} \, j^\gamma_{{\bf k}
s f}\,( v ^o _{\bf k} ) ^\ast
\, \vec{v}_{\bf k} \cdot \vec{e}_h \,
 F({\bf k},\omega)\, ,
\end{equation}
which depends on  the superconducting gap
function  not only through
its magnitude but also through
$\Delta ^o_{\bf k}$ itselfe.
Due to the additional triplet part however,
the result is still gauge invariant.
The function $F({\bf k},\omega)$ correspond
to the  $ G_{1(2)}({\bf k},\omega)$
for $\chi^{(h)}_{\alpha \beta \gamma}( \omega)$.
Considering a magnetic field parallel to the
x-axis (in plane), the nonvanishing
elements of $\chi ^{(h)}_{\alpha \beta \gamma}
( \omega)$ are:
$\alpha \beta \gamma \in
\{yyy,xxy,xyx,yxx,zzy,zyz,yzz\}$.
This results from the combination of the
transformation properties of the
normal state matrix elements and of the
symmetry sensitive term
$(v ^o _{\bf k}) ^\ast \vec{v}_{\bf k}
\cdot \vec{e}_h$.
Using Eq.(~\ref{vmatrix}),
the ${\bf k}$-dependence of the latter
results  from
the  irreducible representations given
in table~\ref{tab1}.

The above matrix elements lead to a rotation
of the polarization of the
incident light due to the interference of
the singlet and triplet
states at the surface of a superconductor.
Thus nonlinear magneto-optics,
unlike linear optical probes, provides
indeed an optical method to
discriminate different superconducting pairing
symmetries by exclusively
employing the effect of optical photons to
low energy excitations.
In the next paragraph, we discuss the
numerical results of the above model
bandstructure for the most realistic
experimental setups
and show how experiments can distinguish
between certain
symmetries of the gap function using
nonlinear magneto-optics.

\section{RESULTS}

In this section we discuss the numerical
 results obtained for the tensor
elements $\chi^{(0)}_{zzz}$ of second
harmonic generation without
magnetic field and  $\chi^{(h)}_{yzz}$
 which gives rise to the
rotation of the polarization plane for
an applied magnetic field
parallel to the $x$-axis.
For simplicity we neglect the dispersion
of the initial and the final
state and consider solely the
${\bf k}$-dependence which results from
the superconducting gap function.
The momentum summations are performed within
the two dimensional
Brillouin zone using  $81 \times 81$
${\bf k}$-points.

The calculations are performed for a magnetic
field of $9 \, {\rm T}$
(corresponding to a field induced band splitting
of 0.5 meV)
and a temperature of $1.5\, {\rm K}$.
The magnitude of the singlet part of the gap
 function is assumed to be
$ 5 \, {\rm meV}$. Furthermore, the magnitude
of the dipole matrix elements
is estimated to be 10$^{-11}$ m.
All results presented here are not sensitive to
the specific set of parameters
chosen, but are typical for reasonable values of
the corresponding energy
scales. This was checked by systematically varying
the dependence of the
nonlinear susceptibility on the magnitude of the
gap function, the magnetic
field, the position of the initial and final states
$E_i$ and $E_f$, the
temperature and the linewidth broadening $\delta$.

In Figs.~\ref{fig1}(a) and (b), we show
$\omega ^2 {\rm Im}\chi^{(o)}_{zzz}(\omega)$ and
$\omega ^2 {\rm Im} \chi^{(h)}_{yzz}(\omega)$
for an isotropic $s$-wave.
For the conventional SHG, we find a line shape
similar to the real part of a
Lorentzian, which is typical for a three level
system discussed in this paper.
More interestingly, the fine structure of the
peak, shown in the inset of
Fig.~\ref{fig1}(a) clearly shows the energy
scale of the superconducting gap.
Comparing this behavior with the Kerr signal
$\omega ^2 {\rm Im} \chi^{(h)}_{yzz}(\omega)$
of  Fig.~\ref{fig1}(b),
one finds that the interference of the singlet
and triplet pairing states
leads to a line shape with several pronounced
zeros and with a fine structure
that yields, besides the energy scale of the
superconducting gap, also
excitations which result from the magnetic
field splitting.
In all our calculations, this line shape was
exclusively observed for an
isotropic $s$-wave and can be considered as
a fingerprint of this symmetry.

In Figs.~\ref{fig2} and ~\ref{fig3}, the
corresponding results for
the anisotropic $s$-wave and the
$d_{x ^2 - y ^2}$-wave are shown.
Although the result for the conventional
SHG is similar to that
of the isotropic $s$-wave, a totally
different line shape of the Kerr
signal results.
This is due to the symmetry dependent
prefactor in $\chi^{(h)}_{yzz}(\omega)$
and can be used to discriminate these
two symmetries from the isotropic
$s$-wave.
Furthermore the fine-structure of these
two symmetries is very different
from the isotropic $s$-wave. Due to the
occurrence of nodes in the gap,
not only a peak but a whole broad band
between $3\, {\rm eV}$ and
$3.01\, {\rm eV}$ is observable. This range
is surprisingly given by twice the
superconducting gap magnitude.
Unfortunately, there are only slight
differences between the two
symmetries shown in Fig.~\ref{fig2}
and ~\ref{fig3}. This is due to the
similar ${\bf k}$-dependence of the triplet
part given in table~\ref{tab1}.
Only the fine structure of the peaks displayed
in the insets of
Fig.~\ref{fig2}(b) and ~\ref{fig3}(b) exhibits
a clear difference, where the
anisotropic $s$-wave has a clear zero at
$3.01\, {\rm eV}$ which is
more or less smeared out for the
$d_{x ^2 - y ^2}$ wave.

In this context it is of importance to
compare these results to the usual
nonlinear Kerr effect in the normal state
 in an external applied field
which can be easily estimated using
previous results by
Pustogowa {\em et al.}~\cite{pustogowa}
 as
\begin{equation}
\frac{\chi^{(h)}_{yzz,normal}}
{\chi^{(0)}_{zzz}}\;\approx\;
\frac{\lambda_{s.o}J(h)}{(\hbar\omega)^{2}}\;.
\end{equation}
Here, the incident photon energy $\hbar\omega$
at resonance is 3 eV, while
the energy splitting $J(h)$ caused by the
external magnetic field $h$ is
0.5 meV (see above). For this small value of $J(h)$,
the formula given in
appendix C of the paper by Pustogowa {\em et al.}
yields a linear dependence.
Thus, we find $\frac{\chi^{(h)}_{yzz,normal}}
{\chi^{(0)}_{yzz}}\;\approx\;
2.7\;\cdot\;10^{-6}$ due to the absence of a
spontaneous magnetization.
{}From this estimate, it follows that the
observability of the new
contribution to the nonlinear Kerr effect
 is clearly guaranteed
for the anisotropic s-wave.
Although, the intensities of the other
symmetries are smaller
than the estimated value of the usual
nonlinear Kerr effect in both
the
normal and superconducting state, we believe
that this effect is
still
observable for the following reason:
Due to the neglect of the dispersion of the
states in our model
bandstructure,
the disapperance of the signal results from
cancellations of
contributions
of the order of magnitude of the anisotropic
 s-wave in the ${\bf
k}$-summation.
This is due to the artificially high symmetry
of this bandstructure.
The simplified description of the high-T$_c$
materials is used
because the aim of this paper is to
demonstrate the strong
interdependence
of the lineshape of the nonlinear magneto-optical
susceptibility and
the
symmetry of the superconducting state.
A more realistic bandstructure immediately
leads to larger
intensities of
$\chi ^{(h)} _{yzz}$, while keeping the
characteristics of the line
shapes of the spectra.

In Fig.~\ref{fig4} we finally show our results
for the $d_{xy}$ symmetry.
Although, this symmetry does not seem to be
the most probable candidate
for the high-T$_c$  materials, it shows most
clearly the symmetry dependence
of the nonlinear magneto-optical Kerr effect.
In contrast to the
anisotropic $s$-wave and the $d_{x ^2 - y ^2}$
wave, there occurs a sign change
between the magnetic and nonmagnetic optical
spectrum.
This is observable since the sign of the Kerr
 spectrum determines the direction
of the rotation of the  polarization axis,
i.e. the Kerr angle.
Furthermore, the satellites shown in the
inset of  Fig.~\ref{fig4}(a) and (b)
cover only the range from $3\, {\rm eV}$
to $3.005 \, {\rm eV}$, i.e.
only one times the gap magnitude.

For the existence of a finite Kerr signal,
it is necessary to break
time reversal symmetry and to apply an external
magnetic field.
This enables one to keep any direction of the
field fixed and to study
the anisotropy of the effects discussed in
this paper.
However, due to the strong but short-ranged
antiferromagnetic correlations
it might also be possible to take advantage
of the locally broken time-reversal
symmetry of the high-T$_c$ materials.
Since a finite Kerr signal is expected for
certain long-range ordered
antiferromagnets~\cite{fr"ohlich}, a
pump-and-probe experiment
(on a time scale faster than the average
lifetime
of the local spin configurations
$\tau_{\rm spin} \approx 10 ^{-12} -
10 ^{-13} \, {\rm s}$)
could be able to resolve the influence
of the neighboring spins on the
site which is excited by the optical
excitation.
Furthermore, for a practical realization
of the experiment, one has to take into
account the possible heating effects of the
incoming light on the sample, which
might lead to a local disappearance of the
 superconducting state.
In view of the comparable excitation energies
of the nonlinear magneto-optical
Kerr effect in ferromagnets, we believe that
 this effect is of minor importance,
since for fs laser pulses sample heating is
of the order of 10 K and thus
negligible. This simple estimate is readily
obtained from the comparison of
laser heating with ns pulses~\cite{vaterlaus}
yielding intensities of
100 MWm$^{-2}$ and temperature rises of the
order of 100 K and typical fs
measurements of the nonlinear
Kerr-effect~\cite{theo} which operate at laser
powers as low as 100 mW focused on spot
diameters of 100 $\mu $m.

In conclusion, we presented a theory for
the nonlinear magneto-optical response
of superconductors. The surface sensitivity
of this experiment is of
particular interest for the simultaneous
occurrence of singlet and triplet pairing
 amplitudes.
Therefore, a suitable material for this
experiment is the
Bi$_2$Sr$_2$CaCu$_2$O$_8$ compound, where
almost no  reconstruction
of the cleaved surface (Bi-O layer)  occurs,
 and where the existence of
the superconducting state in the upper
 CuO$_2$ layer
was  clearly shown in  photoemission
experiments.~\cite{SDW93}.
Since, due to difficulties in the  manufacture
of the tunneling contacts,
all corner junction experiments where so far
 performed with
YBa$_2$Cu$_3$O$_{7-\delta}$ samples, this
 gives also  insight into the
{\em symmetry} of another class of cuprate
compounds.
Note that furthermore  the surface sensitivity
does not depend on the
actual depth beneath the
surface where superconductivity starts,
since the electronic symmetry
of the superconducting state surface is
of relevance in this context
which may or may not be perfectly identical
 to the physical surface.
On the other hand the new nonlinear
magneto-optical effect proposed in
this paper may also be of considerable
importance for interfaces between
$s$- and $d$-wave superconductors.
Furthermore, we performed a group-theoretical
classification of the irreducible
representations of the gap function at the
surface of a tetragonal system.
Based on this theory, we showed that the
interference of singlet and triplet
pairing states in a magnetic field leads to
a new contribution to the
nonlinear magneto-optical Kerr signal which
is sensitive to certain
{\em symmetries} of the superconducting order
 parameter rather than only to its
{\em magnitude}.
This enables us to give the basic line shapes
 of the corresponding optical
spectra and to show that it is possible to
discriminate an isotropic $s$-wave
and a $d_{xy}$ wave from the anisotropic
$s_{x ^2 + y ^2}$ and
$d_{x ^2 - y ^2}$ waves. Unfortunately, it
is not possible to discriminate
between the two latter symmetries which seem
to be the most probable
symmetries of the high-$T_c$ materials.
Nevertheless, we believe that the nonlinear
optic can yield information
complementary to the  tunneling experiments
and might be also of importance
in view of the application on  heavy fermion
 superconductors,  where it was
also manifested that the superconducting state
is anomalous~\cite{B84,B90,K92}.

\acknowledgments
We would like to thank Prof. K.-H. Bennemann
for stimulating discussions and his continued
interest in this work.

\begin{table}
\begin{tabular}{|c||c|c|}
\hline
irred. representation & singlet part &
 triplet part \\ \hline  \hline
${\cal A}_1$ & $const.$ ,
$\cos(k_x)+\cos(k_y)$ &
$\vec{e}_x \sin(k_y) - \vec{e}_y \sin(k_x)$
 \\  \hline
${\cal A}_2$ &
$(\cos(k_x)-\cos(k_y))\sin(k_x) \sin(k_y)$ &
$\vec{e}_x \sin(k_x) + \vec{e}_y \sin(k_y)$
\\  \hline
${\cal B}_1$ &  $\cos(k_x)-\cos(k_y)$ &
$\vec{e}_x \sin(k_y) + \vec{e}_y \sin(k_x)$
 \\  \hline
${\cal B}_2$ &  $ \sin(k_x) \sin(k_y)$ &
$\vec{e}_x \sin(k_x) - \vec{e}_y \sin(k_y)$
\\  \hline
${\cal E}$ &  $ \sin(k_x) \sin(k_z)$,
$\sin(k_y) \sin(k_z)$
& $\vec{e}_z \sin(k_x)  ,
 \vec{e}_z \sin(k_y) $ \\  \hline
\end{tabular}
\caption{Singlet and triplet parts of
the gap function   for
the irreducible representations of the
point group $C_4$. $k_x$ and $k_y$ are
the components of the in plane
momentum. $k_z$ represents an additional
quantum number, which changes
sign if one interchanges the two  paired
electrons and which is related to the
layer index at the surface.}
\label{tab1}
\end{table}

\begin{figure}
\caption{$\omega ^2 {\rm Im}\chi^{(o)}_{zzz }
(\omega)$ (a) and
$\omega ^2 {\rm Im} \chi^{(h)}_{yzz}(\omega)$
(b) for an isotropic $s$-wave
symmetry of the gap function. Note the very
different line shapes of
the two experiments, typical for the
isotropic $s$-wave.
The insets show the fine structure of the
results near the excitation energy
of $3\, {\rm eV}$. For better visibility
the spectra are artificially
broadened by a Lorentzian width $\delta $
of 50 meV in the main figures and of
0.5 meV in the insets throughout.
Consequently, the inset peak heights
differ from those of the main figures.}
\label{fig1}
\end{figure}
\begin{figure}
\caption{As in Fig. 1, but for an
anisotropic $s_{x ^2 + y ^2}$ wave.
Note the difference of the magnetic line
shape compared to the isotropic
$s$-wave and the broad fine structure up
to two times the magnitude of the gap,
shown in the inset.}
\label{fig2}
\end{figure}
\begin{figure}
\caption{As in Fig. 1, but for a
$d_{x ^2 - y ^2}$ wave.
The spectra are very similar to the
$s_{x ^2 + y ^2}$ wave, which results from
the similar ${\bf k}$-dependence of
the triplet pairing amplitude, given in the
table. Slight differences occur for
the fine structure near $3\, {\rm eV}$.}
\label{fig3}
\end{figure}
\begin{figure}
\caption{As in Fig. 1, but for a
$d_{x  y }$ wave.
In distinction to the $s_{x ^2 + y ^2}$
and $d_{x ^2 - y ^2}$ wave, the signs
of the nonmagnetic and magnetic spectra
are different. Furthermore the
satellites shown in the inset occur only
up to the magnitude of the gap
function.}
\label{fig4}
\end{figure}
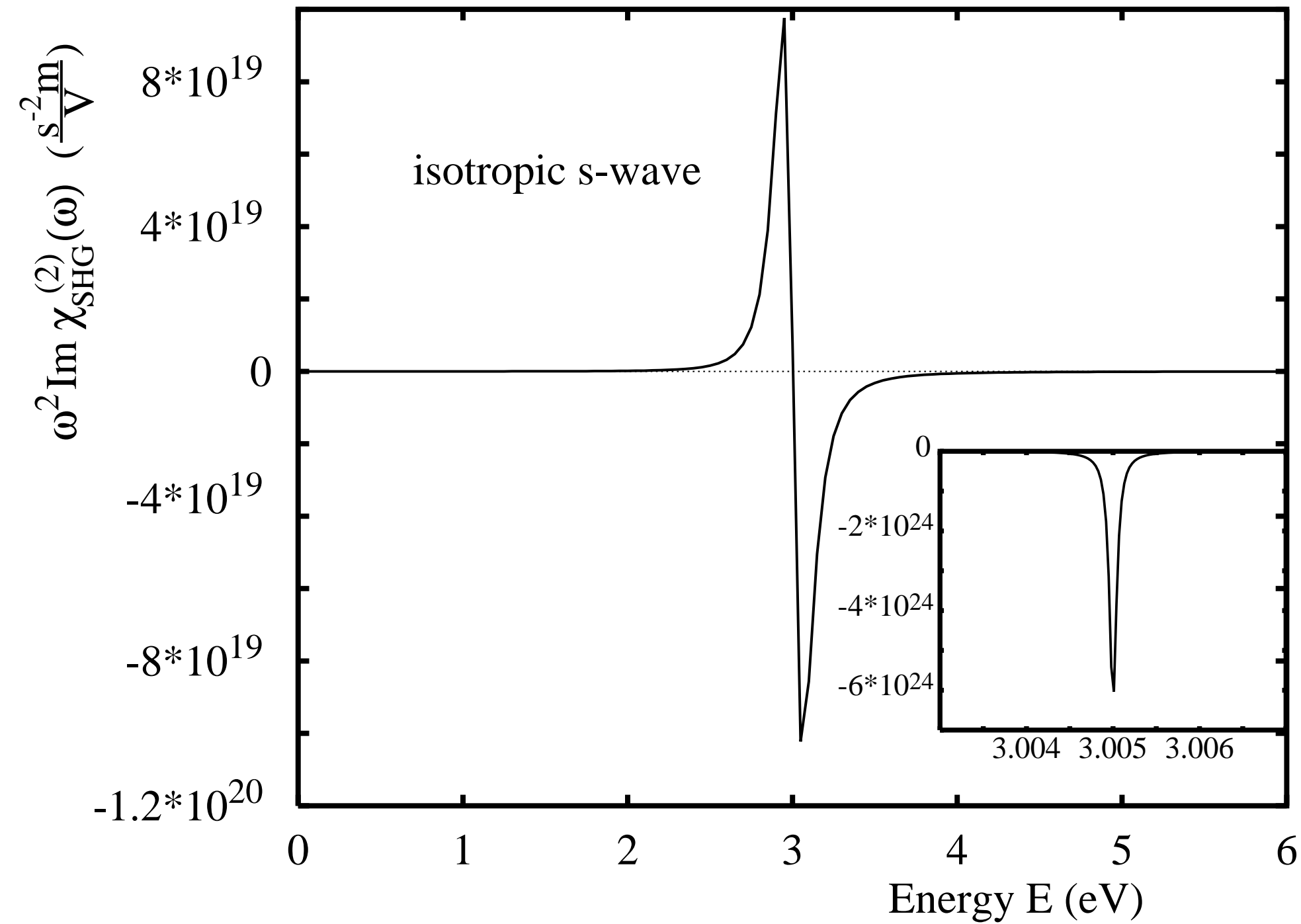
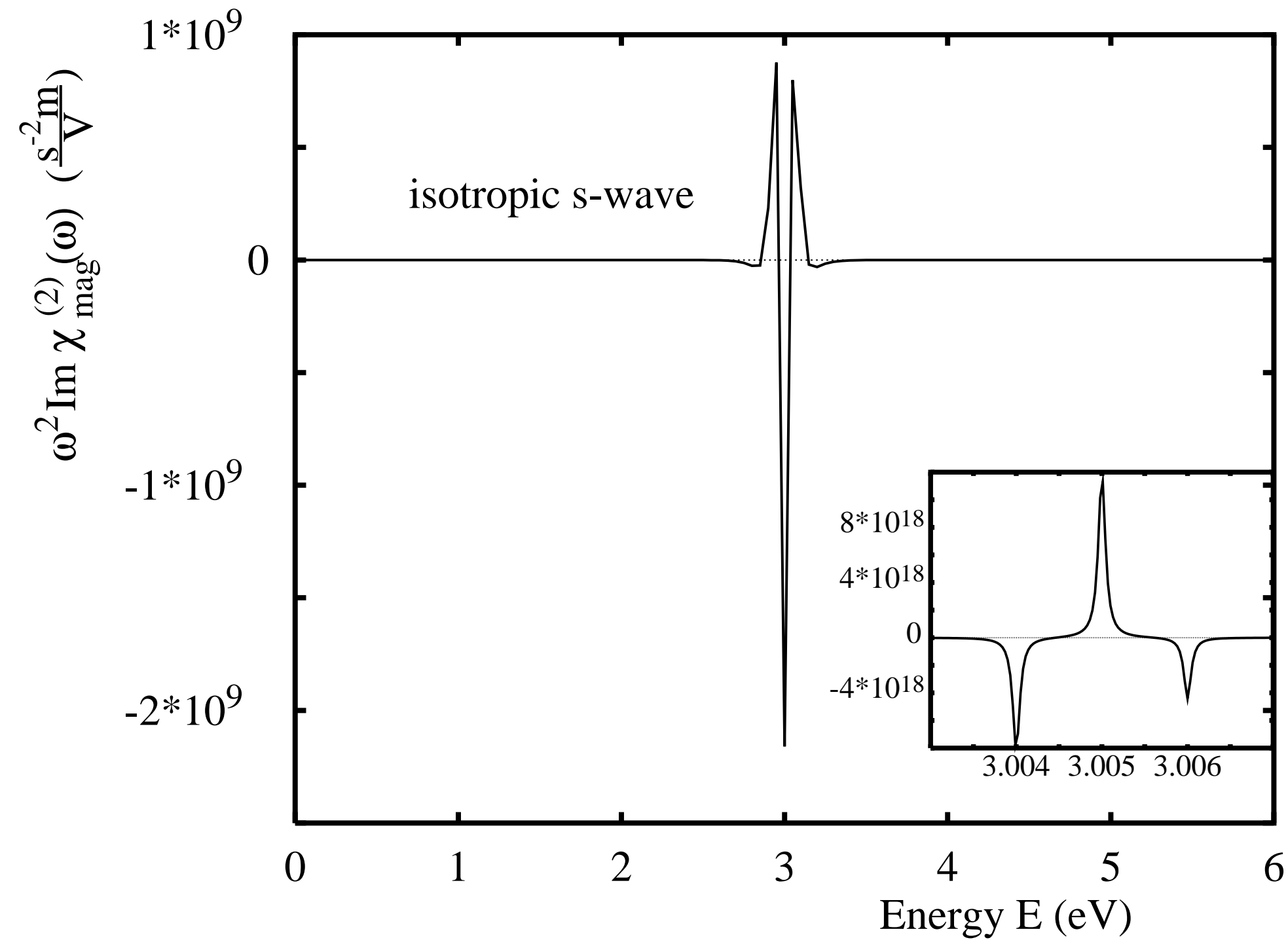
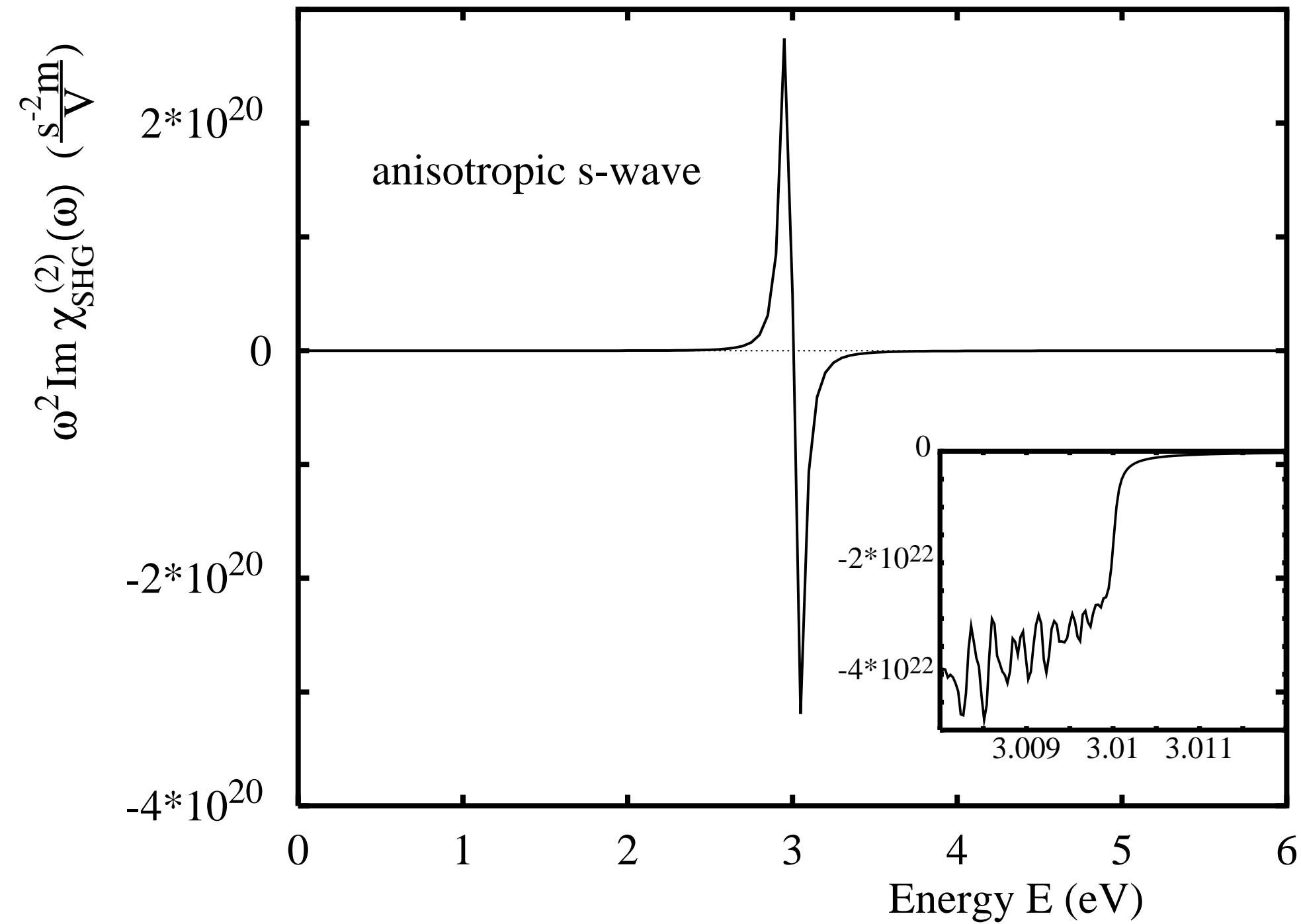
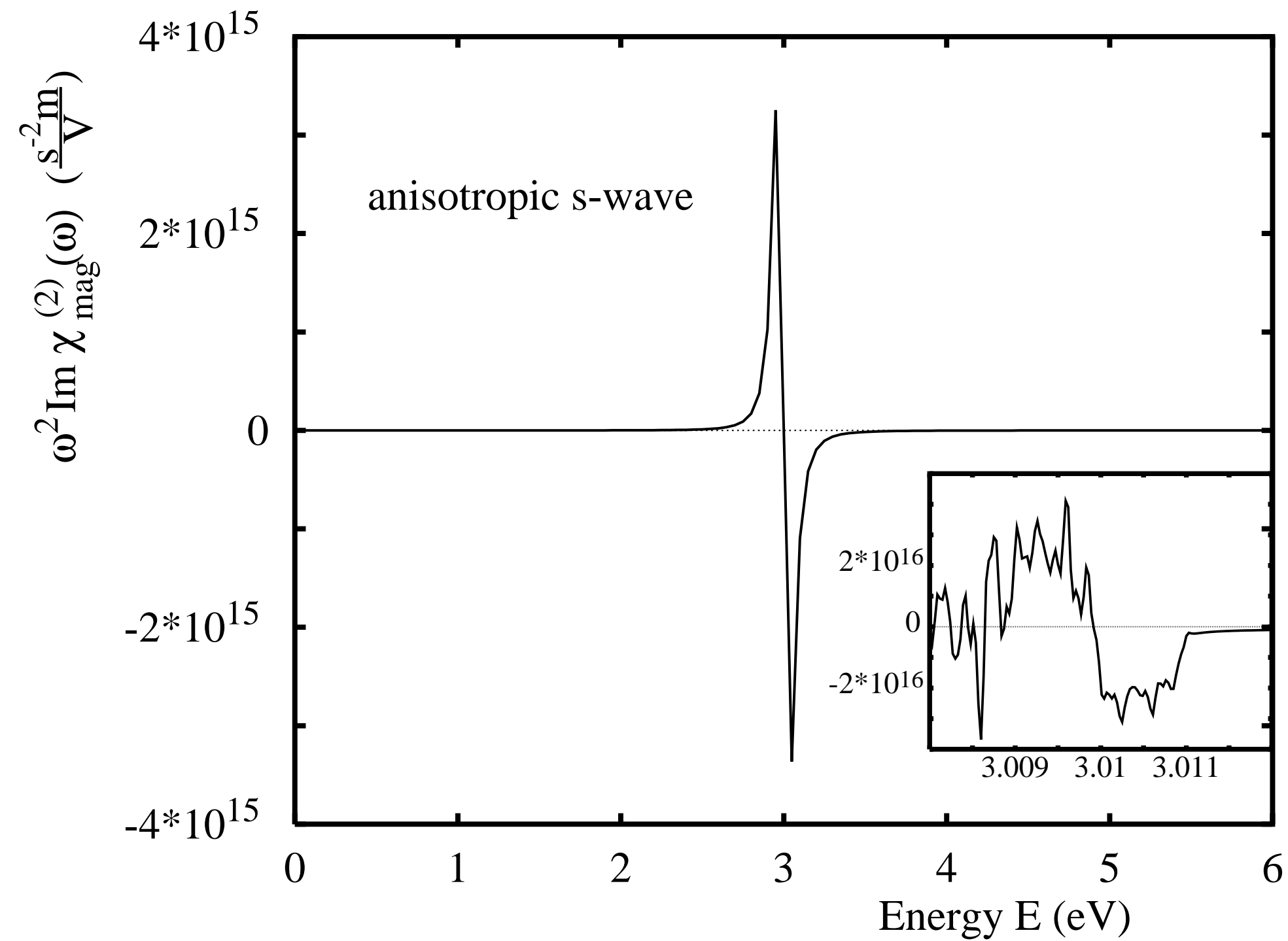
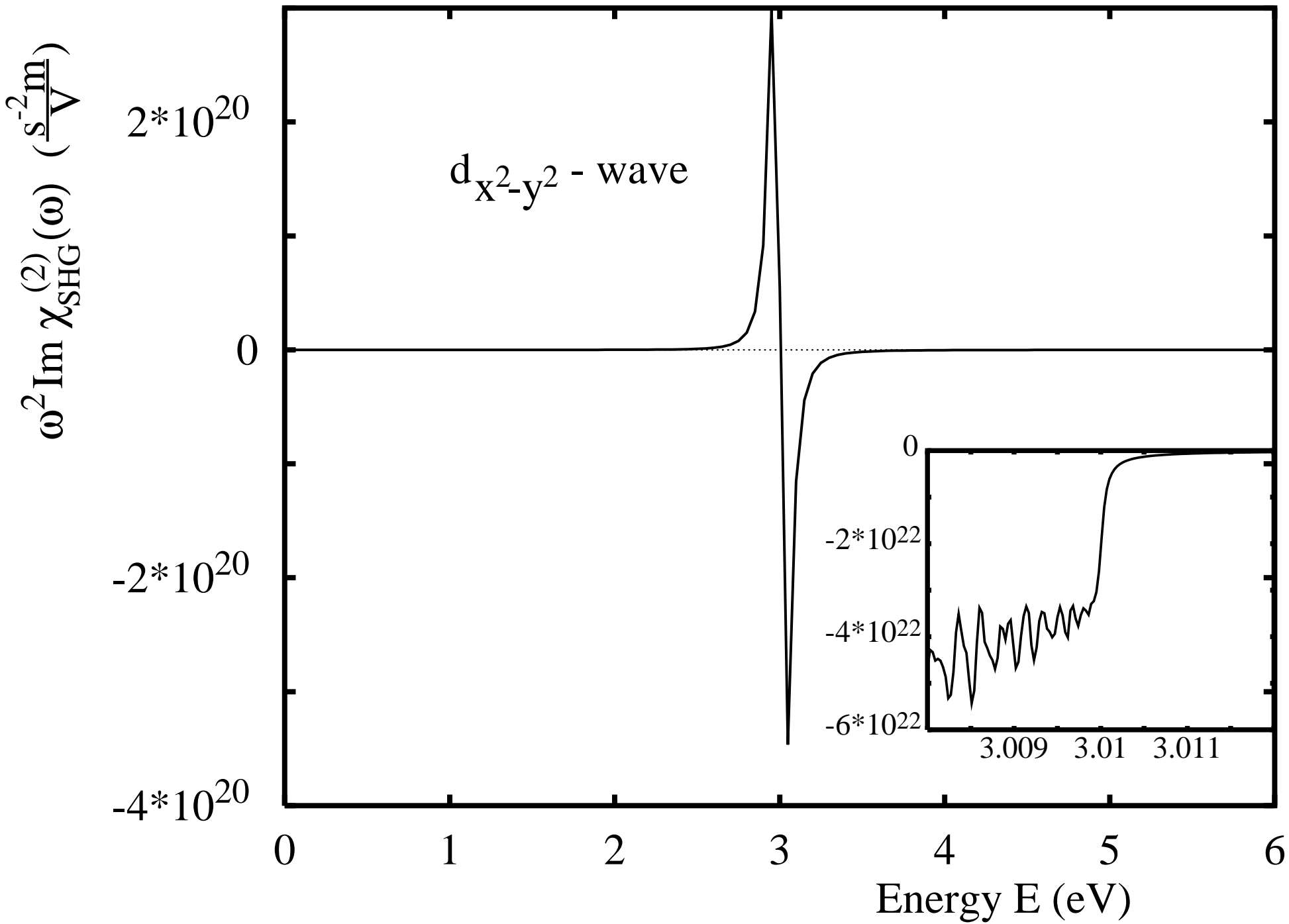
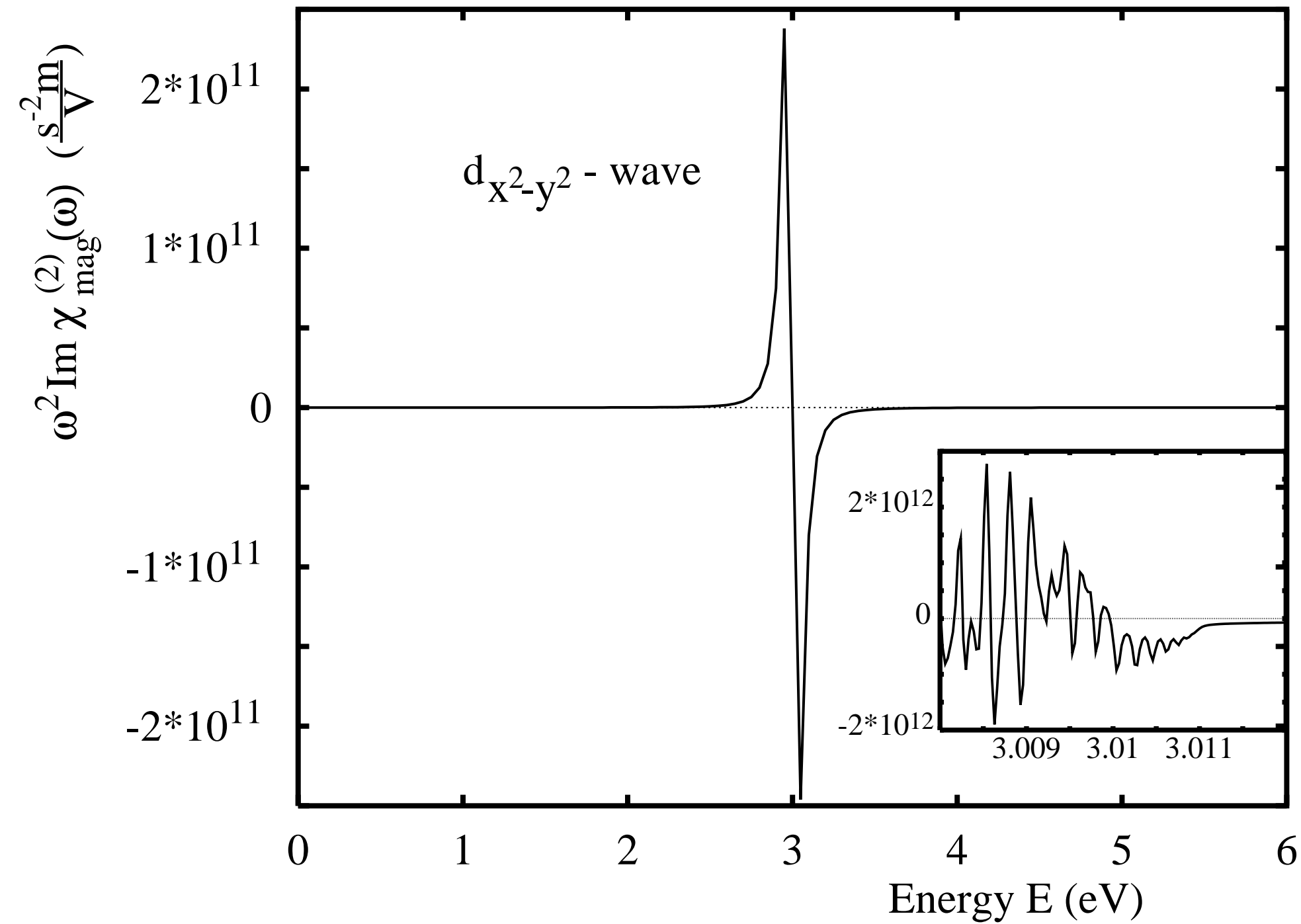
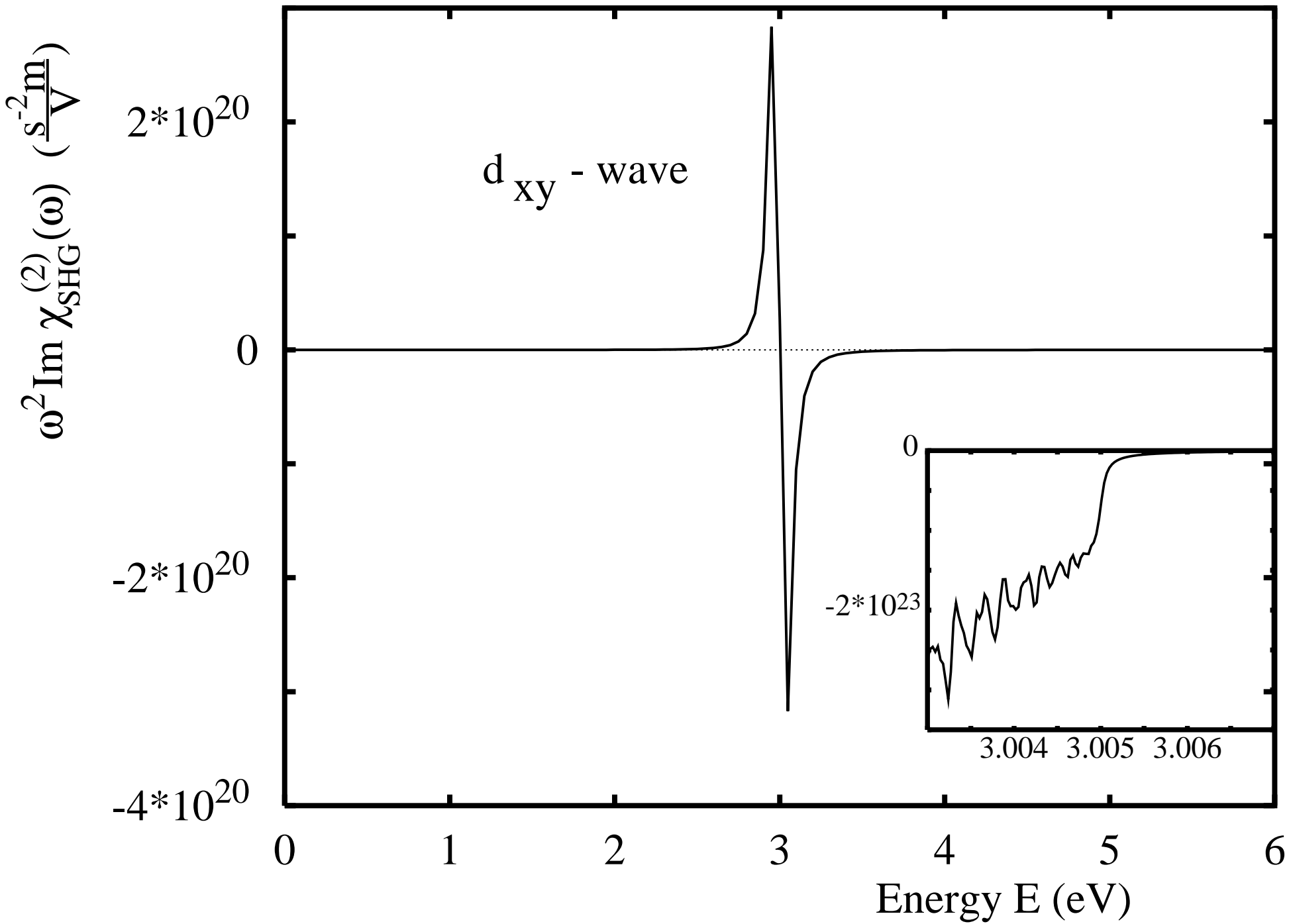
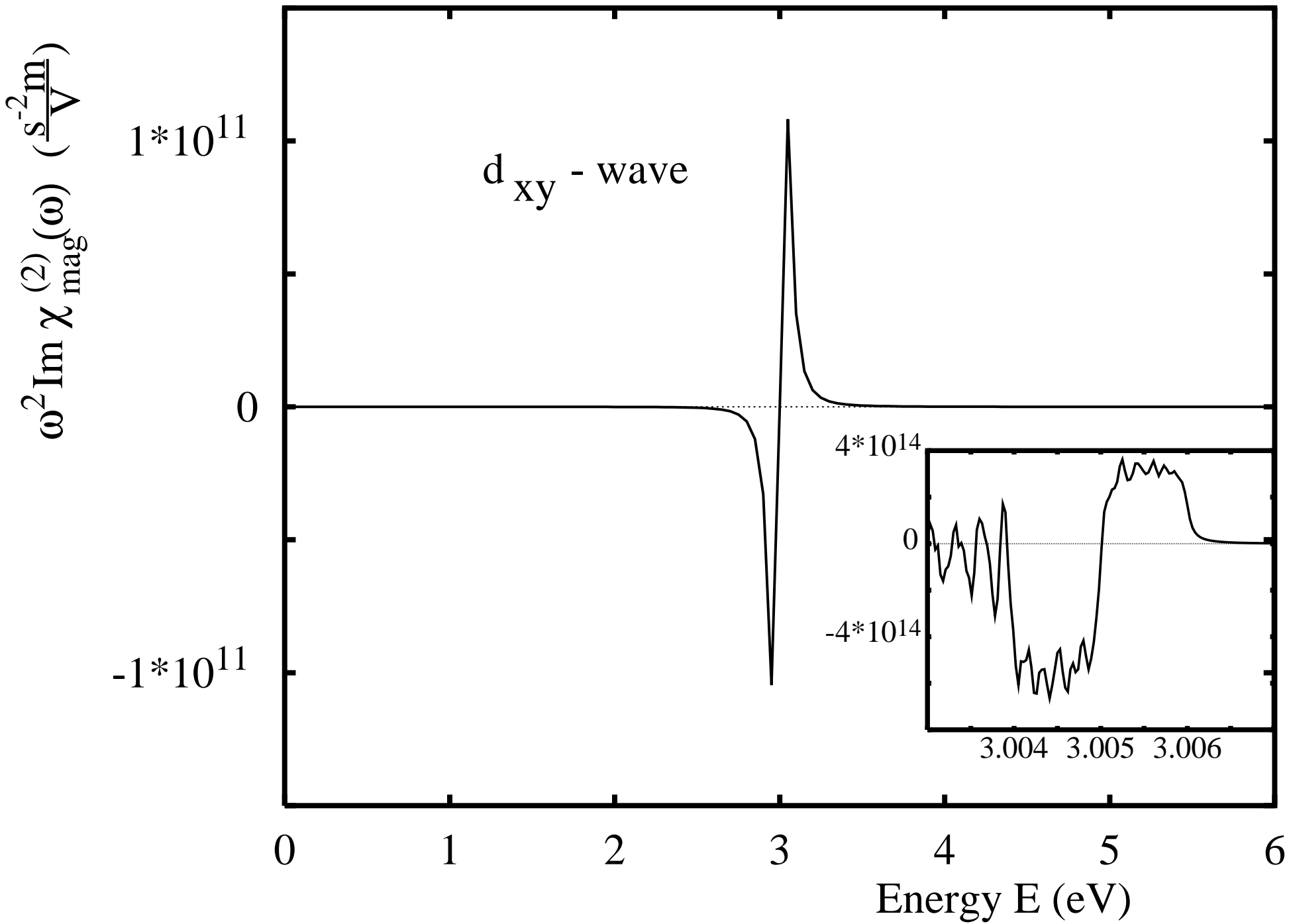
\end{document}